\documentclass{article}
\usepackage{epsfig}
\usepackage{color}
\usepackage{amsmath}
\usepackage{amssymb}

\makeatletter
\@addtoreset{equation}{section}

\makeatother

\begin{document}

\begin{flushright}
March 2012

SNUTP12-001
\end{flushright}

\begin{center}

\vspace{5cm}

{\LARGE 
\begin{center}
Supersymmetry Breaking in Chern-Simons-matter Theories
\end{center}
}

\vspace{2cm}

Takao Suyama \footnote{e-mail address : suyama@phya.snu.ac.kr}

\vspace{1cm}

{\it 
BK-21 Frontier Research Physics Division

and 

Center for Theoretical Physics, 

\vspace{2mm}

Seoul National University, 

Seoul 151-747 Korea}

\vspace{3cm}

{\bf Abstract} 

\end{center}

Some of supersymmetric Chern-Simons theories are known to exhibit supersymmetry breaking when the Chern-Simons level is less than a certain number. 
The mechanism of the supersymmetry breaking is, however, not clear from the field theory viewpoint. 
In this paper, we discuss vacuum states of ${\cal N}=2$ pure Chern-Simons theory and ${\cal N}=2$ Chern-Simons-matter theories of quiver type using related 
theories in which Chern-Simons terms are replaced with (anti-)fundamental chiral multiplets. 
In the latter theories, supersymmetry breaking can be shown to occur by examining that the vacuum energy is non-zero. 

\newpage

\vspace{1cm}

\section{Introduction}

\vspace{5mm}

It has turned out that supersymmetric Chern-Simons-matter theories play a very important role in the study of M-theory. 
The low energy dynamics of M2-branes is described by such a three-dimensional field theory. 
Details of the theory, the amount of supercharges and matter contents for 
example, are determined by the eleven-dimensional space-time in which the M2-branes live. 
M2-branes living in the flat space-time is expected,  at least when 
the number of the M2-branes are two, to be described by BLG theory \cite{Bagger:2006sk}-\cite{Gustavsson:2007vu} 
which can be written as a Chern-Simons-matter theory \cite{VanRaamsdonk:2008ft}. 
ABJM theory \cite{Aharony:2008ug} provides a description of an arbitrary number of M2-branes at the tip of an orbifold $\mathbb{C}^4/\mathbb{Z}_k$. 
This has ${\cal N}=6$ supersymmetry and ${\rm U}(N)\times{\rm U}(N)$ gauge group with the Chern-Simons levels $(+k,-k)$. 

The ${\cal N}=6$ supersymmetry of the ABJM theory is preserved even when the gauge group is replaced with ${\rm U}(N_1)\times {\rm U}(N_2)$ 
\cite{Hosomichi:2008jb}\cite{Aharony:2008gk}, at least at the classical level. 
However, it is argued \cite{Aharony:2008gk} that this theory does not exist as a physically sensible theory when 
\begin{equation}
|N_1-N_2|>k
    \label{ABJ}
\end{equation}
is satisfied. 
This is based on the following observation \cite{Aharony:2008gk}. 
When some of the bi-fundamental fields in the theory have non-vanishing vevs, at low energy, 
the theory with $N_1\ne N_2$ is reduced to an effective theory including a decoupled 
${\cal N}=3$ pure Chern-Simons theory with the gauge group ${\rm SU}(|N_1-N_2|)$ and the level $k$. 
The vevs correspond to a point in the moduli space of the ${\cal N}=6$ theory in which no classical vacua are lifted by quantum corrections. 
However, it is known \cite{Bergman:1999na}\cite{Ohta:1999iv} that the supersymmetry is spontaneously broken in ${\cal N}=3$ pure Chern-Simons theory 
when the parameters satisfy the condition (\ref{ABJ}). 
This apparent contradiction would come from the assumption that the ${\cal N}=6$ theory satisfying the condition (\ref{ABJ}) exists. 

\begin{figure}[tbp]
\begin{center}
\includegraphics{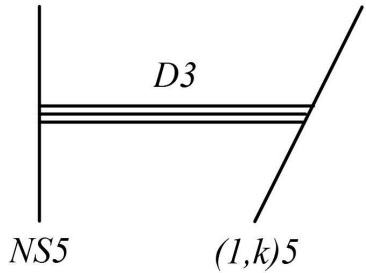}
\caption{The brane configuration realizing ${\cal N}=2$ pure Chern-Simons theory. }
   \label{brane1}
\end{center}
\end{figure}

Supersymmetry breaking in a supersymmetric Chern-Simons theory can be understood most easily when the theory can be realized on some D-branes in 
string theory. 
For example, ${\cal N}=2$ pure Chern-Simons theory with the gauge group ${\rm SU}(N)$ and the level $k$ is realized on D3-brane segments suspended between 
an NS5-brane and a $(1,k)$5-brane \cite{Kitao:1998mf}\cite{Bergman:1999na}, as depicted in Figure \ref{brane1}. 
It is known that such a brane system obeys the s-rule \cite{Hanany:1996ie} by which one can check whether the supersymmetry is broken or not. 
For the ${\cal N}=2$ pure Chern-Simons theory, it turns out that the supersymmetry is broken completely if and only if 
\begin{equation}
N>k
    \label{SUSYbreaking}
\end{equation}
is satisfied \cite{Bergman:1999na}\cite{Ohta:1999iv}. 
This condition can be phrased as follows: 
the supersymmetry is broken when the 't Hooft coupling $\lambda=N/k$ is larger than 1. 
It is usually the case that a gravity dual, assuming it exists, will be classical and easy to be handled when the 't Hooft coupling, as well as $N$, is large. 
Therefore, the above condition (\ref{SUSYbreaking}) implies that, even if there exists a classical gravity dual for ${\cal N}=2$ pure Chern-Simons theory, 
the supersymmetry should be also broken spontaneously in the gravity side. 
This kind of argument can apply to other Chern-Simons-matter theories. 
If one theory exhibits a pattern of supersymmetry breaking, then the information can be used to discuss the expected gravity dual, whose discussion may include whether 
such a dual should exist or not. 
If there is no known brane construction for a Chern-Simons-matter theory which one would like to consider, one may try to generalize the arguments in 
\cite{Witten:1999ds}\cite{Ohta:1999iv} to the theory. 
However, it seems that such generalizations are not always so straightforward technically. 
Supersymmetry breaking in Chern-Simons-matter theories was discussed recently in \cite{Niarchos:2009aa}\cite{Morita:2011cs}\cite{Smilga:2012uy}. 
A Hamiltonian formulation was recently developed in \cite{Agarwal:2012bn} for supersymmetric Yang-Mills-Chern-Simons theories. 

In this paper, we present a field-theoretic argument to study the supersymmetry breaking in Chern-Simons-matter theories. 
It is different from the one given in \cite{Witten:1999ds}\cite{Ohta:1999iv} 
for pure (Yang-Mills)-Chern-Simons theories. 
In fact, it is not clear in \cite{Witten:1999ds}\cite{Ohta:1999iv} how the supersymmetry is broken. 
We will show that, by replacing the Chern-Simons terms with massive fundamental multiplets using the phenomenon found in \cite{Niemi:1983rq}\cite{Redlich:1983dv}, 
the supersymmetry breaking can be studied 
through investigating whether the scalar potential is always positive or not, a quite familiar criterion to show the supersymmetry breaking. 
We discuss ${\cal N}=2$ pure Chern-Simons theory and a family of quiver-type theories, but our argument can be generalized to the other theories, as long as 
the quantum effects in corresponding theories are well understood. 
Our argument can show the occurrence of the supersymmetry breaking in a range of parameters. 
For example, for ${\cal N}=2$ pure Chern-Simons theory, we will show that the supersymmetry is broken completely if $N>k+1$, while for the other cases we will 
give a plausible argument which is consistent with the pattern derived based on the analysis of the D-brane system \cite{Bergman:1999na}\cite{Ohta:1999iv}. 

The basic idea in this paper is in common with the analysis based on brane configurations. 
That is, if one would like to know about the low energy physics of a theory, then one may instead investigate any desired theory as long as it shares the same IR 
physics with the former. 
The worldvolume theory for the brane configuration (Figure \ref{brane1}) is one choice of the UV theory. 
An advantage of our choice of the UV theory is that it always exists. 

This paper is organized as follows. 
Section \ref{review} reviews necessary facts about three-dimensional gauge theories. 
${\cal N}=2$ pure Chern-Simons theory is analyzed in section \ref{pureCS}. 
Section \ref{quiver type} 
analyzes ${\cal N}=2$ Chern-Simons theory with gauge group ${\rm SU}(N_1)\times{\rm SU}(N_2)$ with the level $(+k,-k)$ coupled to bi-fundamental matters.

\vspace{1cm}

\section{Real masses and induced Chern-Simons term} \label{review}

\vspace{5mm}

There is a close similarity between ${\cal N}=2$ supersymmetric gauge theories in three dimensions and 
${\cal N}=1$ supersymmetric gauge theories in four dimensions. 
Many of the former theories are obtained from the latter ones through the dimensional reduction. 
Some properties, for example the existence of the non-renormalization theorem for F-terms, inherit from four dimensions to three dimensions. 
Even some non-perturbative phenomena occur in quite similar manner in four and three dimensions. 
There are, of course, some properties which are specific to three-dimensional theories. 
One example is the presence of the real mass terms, and another is the presence of Chern-Simons terms. 
Interestingly, they have a close connection, as will be reviewed in this section. 
Some basic facts about three-dimensional supersymmetric theories can be found, for example, in \cite{Aharony:1997bx}. 

\vspace{5mm}

The mass terms which are familiar in the four-dimensional theories are the complex mass terms which appear as quadratic terms in the superpotential. 
In three dimensions, there is another way to give masses to matter fields while preserving ${\cal N}=2$ supersymmetry. 
Such mass terms are known as real mass terms. 

The origin of the real mass terms can be easily deduced from the following observation. 
The kinetic term of a scalar field $\Phi$ coupled to a gauge field $A_\mu$ in four dimensions provides, through the dimensional reduction, 
\begin{equation}
| D_\mu\Phi |^2 \hspace{5mm} \rightarrow \hspace{5mm} | D_{\hat{\mu}}\Phi |^2 + | A_3\Phi |^2, 
\end{equation}
where $\hat{\mu}=0,1,2$. 
In three dimensions, $A_3$ is a scalar. 
If $A_3$ has a non-zero vev, some components of $\Phi$ become massive. 

This suggests that a mass term can be introduced as a background vector multiplet $V_m$ in which only the third component of the gauge field, with respect to the 
four-dimensional viewpoint, has a non-zero vev, 
\begin{equation}
V_m = \theta\bar{\theta}m. 
\end{equation}
This indeed gives a mass term as follows, 
\begin{equation}
\int d^4\theta\ \Phi^\dag e^{V_m}\Phi = \int d^4\theta\ \Phi^\dag\Phi + m^2|\phi|^2+m\bar{\psi}\psi. 
\end{equation}
Since $V_m$ is real, the mass parameter $m$ must be real. 

In three dimensions, there is no chiral anomaly. 
Therefore, the charge assignment of matter multiplets which couples to $V_m$ is less restrictive. 

\vspace{5mm}

\begin{figure}[tbp]
\begin{center}
\includegraphics{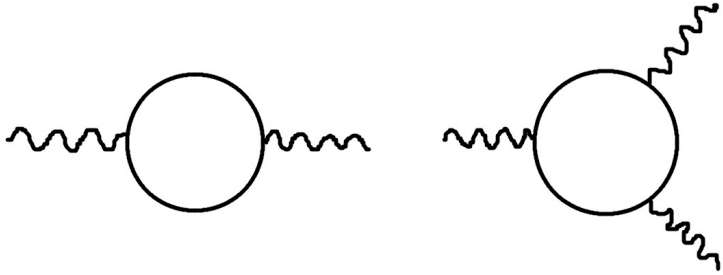}
\caption{The 1-loop diagrams which induce Chern-Simons term. The wavy lines are gauge fields and the solid lines are fermions. The left diagram induces the term with 
a derivative, and the right diagram induces the trivalent vertex of the gauge fields. }
   \label{1-loop}
\end{center}
\end{figure}

Suppose that the mass parameter $m$ is very large. 
Naively, one might expect that the chiral multiplet $\Phi$ can be simply integrated out. 
However, there must be a remnant of $\Phi$ since the original theory with a non-zero $m$ may break the parity invariance \cite{Deser:1981wh}, 
and the low energy effective theory must remember it. 
The remnant is nothing but a Chern-Simons term \cite{Niemi:1983rq}\cite{Redlich:1983dv}. 
It is induced from one-loop diagrams depicted in Figure \ref{1-loop}. 
Contributions from the other diagrams are suppressed by $m^{-1}$ which is negligible at low energy. 
The Chern-Simons level is determined by the representation of the matter fermions circulating the loop. 
If there is a fermion in the representation $R$, the level $k$ is given as 
\begin{equation}
k = \frac12\mbox{sgn}(m)C_2(R), 
\end{equation}
where $C_2(R)$ is the second Casimir for $R$. 
The representation $R$ may be reducible, implying that the level induced by multiple fermions is the sum of contributions from each fermions. 
For the gauge group ${\rm SU}(N)$, $C_2(R)$ is normalized such that it is 1 for the fundamental representation. 

If there is only one massive fundamental chiral multiplet, then it induces a Chern-Simons term with a fractional level, and the gauge symmetry turns out to be 
broken. 
Suppose that there is a pair $(\Phi, \widetilde{\Phi})$ of chiral multiplets in the representations $(R,\bar{R})$ of a gauge group whose real masses are given as follows, 
\begin{equation}
\int d^4\theta \left[ \Phi^\dag e^{V+\theta\bar{\theta}m_1}\Phi +  \widetilde{\Phi}e^{-V-\theta\bar{\theta}m_2}\widetilde{\Phi}^\dag \right]. 
\end{equation}
If both $\Phi$ and $\widetilde{\Phi}$ are integrated out, a Chern-Simons term is induced with the level 
\begin{equation}
k = \frac12[ \mbox{sgn}(m_1)-\mbox{sgn}(m_2) ]C_2(R). 
\end{equation}
Therefore, if $\mbox{sgn}(m_1m_2)>0$, then no Chern-Simons term is induced, while otherwise $k$ is an integer. 
In both cases, the gauge symmetry is preserved. 
It would be convenient to define a vector mass $m_v$ and an axial mass $m_a$ as 
\begin{equation}
m_v = \frac12(m_1+m_2), \hspace{5mm} m_a = \frac12(m_1-m_2). 
\end{equation}
A Chern-Simons term is induced only if $m_a$ is non-zero. 
If the gauge group has a ${\rm U}(1)$ factor, then $m_v$ can be absorbed by shifting the scalar field in the ${\rm U}(1)$ vector multiplet. 
The effects of the vector mass on the low energy dynamics is studied in \cite{deBoer:1997kr}\cite{Aharony:1997bx}.

\vspace{1cm}

\section{Pure Chern-Simons theory} \label{pureCS}

\vspace{5mm}

In this section, we consider vacuum states of ${\cal N}=2$ pure Chern-Simons theory. 
For definiteness, we specify the gauge group to be SU$(N)$ with the Chern-Simons level $k$. 
It is known \cite{Bergman:1999na}\cite{Ohta:1999iv} that this theory has supersymmetric vacua when $N\le k$, 
which can be shown based on the analysis of the corresponding brane configuration, 
while otherwise the supersymmetry is completely broken. 
Exactly speaking, \cite{Bergman:1999na}\cite{Ohta:1999iv} discussed the vacuum states of ${\cal N}=2$ Yang-Mills-Chern-Simons theory which appears as the 
worldvolume theory on D3-branes in the brane configuration. 
However, the properties of the vacuum states of this theory should be the same as those of ${\cal N}=2$ 
pure Chern-Simons theory since the Yang-Mills term is 
irrelevant in three dimensions, and therefore it can be ignored at low energy. 

The breakdown of the supersymmetry is rather difficult to show purely from the field theory viewpoint. 
For example, it cannot be shown by examining the presence of a non-vanishing vacuum energy. 
A rather indirect field theoretic argument was given in \cite{Witten:1999ds} for ${\cal N}=1$ Yang-Mills-Chern-Simons theory, and its extension to the 
${\cal N}=2,3$ theories was discussed in \cite{Ohta:1999iv}. 
In these analyses, the presence of the Yang-Mills term seems to be crucial. 

In the following, we argue the occurrence of supersymmetry breaking in a field theory which flows to ${\cal N}=2$ Yang-Mills-Chern-Simons theory, 
and therefore to ${\cal N}=2$ pure Chern-Simons theory, in the IR limit. 
As discussed previously, by definition, the UV theory has vacuum states which share the same properties with those 
of ${\cal N}=2$ pure Chern-Simons theory. 
It will turn out below that the analysis of the former would be more transparent than the analysis of the latter. 
It should be noted that our analysis has also a disadvantage: our analysis can show the occurrence of supersymmetry breaking for $N-1>k$ but for the other cases 
we cannot draw any definite conclusion. 
It is, however, possible to argued that the other cases also seem to be compatible with the known pattern of supersymmetry breaking. 

\vspace{5mm}

The theory which will be considered in this section is 
${\cal N}=2$ gauge theory in three dimensions with the gauge group SU$(N)$ coupled to $k$ chiral multiplets $Q^I$ $(I=1,2,\cdots,k)$ 
in the fundamental representation of ${\rm SU}(N)$ and $k$ chiral multiplets $\widetilde{Q}^I$ in the anti-fundamental representation. 
All the chiral multiplets have a common axial mass $m$. 
The Lagrangian is given as follows, 
\begin{equation}
L = \int d^4\theta \left[ Q_Ie^{V+\theta\bar{\theta}m}Q^I+\widetilde{Q}^Ie^{-V+\theta\bar{\theta}m}\widetilde{Q}_I \right] 
     + \left[ \int d^2\theta \ \mbox{Tr}\ W^\alpha W_\alpha + \mbox{\rm h.c.} \right], 
    \label{SQCD}
\end{equation}
where $Q_I=(Q^I)^\dag$ and $\widetilde{Q}_I=(\widetilde{Q}^I)^\dag$. 
If $m$ is zero, this theory is nothing but the dimensional reduction of the four-dimensional supersymmetric QCD \cite{Intriligator:1995au} with $N_c=N$ and $N_f=k$. 

If $m$ is large, or if one considers this theory at an energy scale much lower than $m$, the chiral multiplets can be integrated out. 
As explained in section \ref{review}, this procedure induces a Chern-Simons term with the level $k$. 
At low energy, the theory (\ref{SQCD}) becomes equivalent to ${\cal N}=2$ Yang-Mills-Chern-Simons theory without matter. 
At a lower energy, the Yang-Mills term becomes irrelevant due to a positive mass dimension of the gauge coupling, and it can be dropped in the IR limit. 
Therefore, the theory (\ref{SQCD}) shares the same low energy properties with ${\cal N}=2$ pure Chern-Simons theory. 

\vspace{5mm}

\subsection{Classical vacua}

\vspace{5mm}

The supersymmetry is completely broken if and only if the scalar potential ${\cal V}$ of (\ref{SQCD}) is always positive. 
At the classical level, since there is no tree-level superpotential, ${\cal V}$ is the sum of the D-term potential $V_D$ and the mass terms including the axial mass,  
\begin{equation}
{\cal V} = V_D+\sum_{I=1}^k \Bigl( \bigl| (\phi+m)q^I \bigr|^2+\bigl| (\phi-m)\widetilde{q}_I \bigr|^2 \Bigr), 
   \label{classicalD}
\end{equation}
where $q^I$ and $\widetilde{q}^I$ are the lowest components of $Q^I$ and $\widetilde{Q}^I$, respectively, and $\phi$ is the adjoint scalar field in the vector multiplet. 

In the case $m=0$, it is well-known that the solutions of ${\cal V}=0$ form a continuous family, the moduli space of vacua. 
If $N>k$, which will be relevant later, the vevs of $q^I$ and $\widetilde{q}^I$ are 
\begin{equation}
q^I = \widetilde{q}_I = \left[ 
\begin{array}{cccc}
a_1 & & & \\
& a_2 & & \\
& & \ddots & \\
& & & a_k \\
& & & 
\end{array}
\right], 
\end{equation}
up to a symmetry transformation. 
The vev of $\phi$ is then required to satisfy  
\begin{equation}
\phi q^I = 0, \hspace{5mm} \phi\widetilde{q}_I = 0. 
    \label{adjoint mass}
\end{equation}
In general, $\phi$ may be non-zero even when $q^I$ and $\widetilde{q}^I$ are non-zero, and the intersection of the Coulomb branch and the Higgs branch may not be 
just a single point. 

The presence of a non-zero $m$ changes the classical moduli space of vacua drastically. 
The vev of $\phi$ can be always diagonalized by a gauge transformation, 
\begin{equation}
\phi = \mbox{diag}(\phi_1, \cdots, \phi_{N}). 
\end{equation}
If some of the eigenvalues of $\phi$ is equal to $\pm m$, then $q^I$ and $\widetilde{q}^I$ can have non-zero vevs. 
By a gauge transformation, those eigenvalues can be arranged such that 
\begin{equation}
\phi_i = \left\{ 
\begin{array}{cc}
+m, & (1\le i\le l_1) \\ -m, & (l_1<i\le l_2) \\ \mbox{other values} & (l_2<i\le N)
\end{array}
\right.
\end{equation}
For the case $l_1=l_2=0$, the vevs of $q^i$ and $\widetilde{q}^I$ must vanish. 
Otherwise, they satisfy 
\begin{equation}
q^I_1 = \cdots = q^I_{l_1} = q^I_{l_2+1} = \cdots = q^I_N = 0, \hspace{5mm} \widetilde{q}^I_{l_1+1} = \cdots = \widetilde{q}^I_N = 0. 
\end{equation}

The classical D-term potential $V_D$ is 
\begin{equation}
V_D = \frac{g_{\rm YM}^2}2\sum_{a=1}^{N^2-1}\left( q_IT^aq^I-\widetilde{q}^IT^a\widetilde{q}_I \right)^2. 
\end{equation}
There is a generator $T$ of ${\rm su}(N)$ whose matrix form in the fundamental representation is 
\begin{equation}
T = \left[ 
\begin{array}{ccc}
-I_{l_1} & & \\
& +I_{l_2} & \\
& & \frac{l_1-l_2}{N-l_2}I_{N-l_2}
\end{array}
\right], 
\end{equation}
where $I_l$ is the $l\times l$ unit matrix. 
The condition $V_D=0$ then implies 
\begin{equation}
\left( q_ITq^I-\widetilde{q}^IT\widetilde{q}_I \right)^2 = \left[ \sum_{I=1}^k\left( \sum_{i=l_1+1}^{l_2} |q^I_i|^2 + \sum_{i=1}^{l_1} |\widetilde{q}^I_i|^2 \right) \right]^2 = 0. 
\end{equation}
This implies that only $q^I=\widetilde{q}_I=0$ is allowed. 

It is well-known that the condition $V_D=0$ can be solved whenever the F-term condition is solved. 
See e.g. \cite{Luty:1995sd}. 
The above calculation shows that the introduction of the axial mass parameter $m$ lifts all the classical Higgs branch. 

\vspace{5mm}

\subsection{Quantum vacua}

\vspace{5mm}

At the quantum level, the scalar potential ${\cal V}$ is modified from the classical one (\ref{classicalD}) by quantum corrections. 
We start our discussion of the quantum vacua with the investigation of the quantum F-term potential. 

\vspace{5mm}

It was shown in \cite{Aharony:1997bx} that a non-perturbative superpotential 
\begin{equation}
W_{\rm np} \propto \left[ Y\det(\widetilde{Q}^IQ^J) \right]^{\frac1{k-N+1}} + \cdots  
    \label{NPW}
\end{equation}
is induced if $N-1>k$, where $Y$ is a chiral superfield related to the adjoint scalar $\phi$ \cite{deBoer:1997kr}\cite{Aharony:1997bx}. 
The dots in (\ref{NPW}) indicate the other terms which do not depend on $Q^I$ nor $\widetilde{Q}^I$. 
The F-term potential derived from 
(\ref{NPW}) exhibits a runaway behavior along the classical Higgs branch, that is, the potential decreases to zero as $\det(\widetilde{Q}^IQ^J)$ increases. 

In the case $m=0$, this theory does not have any stable vacuum state. 
This fact alone does not allow one to conclude that the supersymmetry is broken in this theory, as was pointed out in \cite{Intriligator:2005aw} 
since there could be a supersymmetric state at infinity. 
To check whether the supersymmetry is really broken or not, one has to examine the behavior of the D-term potential at infinity in the classical Higgs branch. 

\vspace{5mm}

Since there is no non-renormalization theorem for the D-terms, it is quite difficult to show the behavior of the quantum-corrected D-term potential explicitly. 
However, it is still possible to argue that the scalar potential 
${\cal V}$ approaches the classical one (\ref{classicalD}) at infinity of the classical Higgs branch at least when $m$ is small. 
The argument given below is an analog of the one in \cite{Affleck:1984xz} for four-dimensional theories. 

Consider first the case $m=0$. 
It was argued in \cite{Witten:1981nf}, 
whose argument can apply to three dimensions, that the quantum-corrected D-term potential $V_D$ does not lift the classical flat directions. 
In fact, this would not be so surprising in three dimensions since the quantum corrections are quite restricted 
simply due to the fact that the gauge coupling constant $g_{\rm YM}$ is dimensionful. 
The mass terms including $\phi$ should also vanish along the classical moduli space since the vacuum configuration satisfy (\ref{adjoint mass}), 
and every $\phi$ in the mass terms must appear in combinations $\phi q^I$ and $\phi\widetilde{q}_I$ due to the gauge invariance. 

When the vevs of $q^I$ and $\widetilde{q}^I$ are large and along the classical Higgs branch, 
some of the vector multiplets which couple to $q^I$ and $\widetilde{q}^I$ become massive and decouple at a very high energy. 
Since the gauge interaction is weak at a high energy scale, the D-term potential has not received large quantum corrections yet. 
Below the energy scale set by the vevs, there remain the vector multiplets with the gauge group ${\rm SU}(N-k)$ and the chiral multiplets $M^{IJ}=\widetilde{Q}^IQ^J$ which 
are decoupled from the vector multiplets. 
Note that a non-generic vevs of $q^I$ and $\widetilde{q}^I$ which may allow a larger gauge group is forbidden by the non-perturbative F-term potential. 
Since the low energy effective theory does not provide a non-trivial D-term potential, the total D-term potential is given approximately by the classical one 
(\ref{classicalD}). 

Now, we introduce a small mass parameter $m$. 
As long as the vevs are large enough, the scalar potential along the classical Higgs branch remains the classical one (\ref{classicalD}). 
As was shown in the previous subsection, the presence of a non-zero $m$ lifts all the classical Higgs branch, implying that a region of 
the classical Higgs branch near infinity is lifted. 
Therefore, there must be a point in the middle of the Higgs branch where the quantum-corrected scalar potential has a global minimum. 
Since the F-term potential derived from (\ref{NPW}) is exact and turns out to be non-zero at the global minimum, 
it is possible to conclude that the supersymmetry is broken completely if 
\begin{equation}
N-1>k. 
\end{equation}

The remaining task is to take $m$ to be large so that the IR theory is equivalent to ${\cal N}=2$ pure Chern-Simons theory. 
We expect that for any finite $m$ there exists a region near infinity of the classical Higgs branch where the D-term potential is approximated by the classical one. 
If this is the case, then the supersymmetry is also broken for the theory with finite but large $m$ which then implies the supersymmetry breaking in ${\cal N}=2$ pure 
Chern-Simons theory. 
It seems natural to expect that the non-zero vacuum energy would increase if $m$ increases, so the supersymmetry breaking seems to be robust against the change of 
$m$. 

\vspace{5mm}

It should be noted that the field $Y$ should be also fixed to realize a stable vacuum state. 
We simply assumed above that those fields would be fixed by the quantum-corrected D-term potential. 
This is because a Chern-Simons term, which is induced by integrating out the matters with 
a non-zero $m$, provides the vector multiplets with a non-zero mass. 
Therefore, the classical flat directions for $\phi$ must be lifted after turning on a non-zero $m$. 
Since the superpotential does not depend on the real mass \cite{Aharony:1997bx}, this lift should be due to a non-trivial D-term. 
This seems to be reasonable from the fact that the Chern-Simons term appears as a D-term. 
For example, the Chern-Simons term for U$(1)$ gauge group is 
\begin{equation}
\int d^4\theta\ \Sigma V, \hspace{5mm} \Sigma = \overline{D}DV. 
\end{equation}
Since the vacuum state corresponds to a point at the middle of the classical Higgs branch, and 
there is no non-renormalization theorem for D-terms for theories with four supercharges, it is quite difficult to explicitly show 
how the field $Y$ would be fixed. 
There are other fields related to $\phi$ \cite{deBoer:1997kr}\cite{Aharony:1997bx} and contributing to the superpotential (\ref{NPW}). 
We also assumed that they are fixed by the D-term potential as for the case of $Y$. 
See also the discussions in \cite{Aharony:1997bx} for the fixing of $\phi$ in the presence of an axial mass. 

\vspace{5mm}

One might worry about a possibility that there could be a discontinuous change of the non-perturbative superpotential between $m=0$ and $m\ne0$. 
If there would be a singularity in the superpotential, it would be due to the appearance of massless particles at $m=0$ which was integrated out in a description for 
$m\ne 0$. 
In the case considered here, the fields which become massless at $m=0$ are $Q$ and $\widetilde{Q}$. 
Since they are retained, there would be no source of singularities in the superpotential in the limit $m\to0$. 
Therefore, the superpotential (\ref{NPW}) should be also valid for $m\ne0$. 

\vspace{5mm}

It is known that the supersymmetry is also broken when $N-1=k$ \cite{Bergman:1999na}\cite{Ohta:1999iv}. 
However, the discussion of this case using the theory (\ref{SQCD}) would not be so simple since there is no superpotential induced in this case \cite{Aharony:1997bx}. 
Therefore, the occurrence of supersymmetry breaking would depend on the details of the D-term potential. 
The following could be a plausible argument suggesting that the supersymmetry is also broken in this case. 
In the case $m=0$, it was shown \cite{Aharony:1997bx} that, although no superpotential is induced, there is a non-trivial constraint on the fields 
\begin{equation}
Y\det(\widetilde{Q}^IQ^J) = 1
   \label{constraint}
\end{equation}
for a suitable choice of units. 
This constraint forbids the origin of the classical Higgs branch to be included in the quantum moduli space of vacua. 
In other words, the vevs of $q^I$ and $\widetilde{q}^I$ cannot be zero at any vacuum. 
After turning on a non-zero $m$, as explained above, the D-term potential grows toward infinity of the classical Higgs branch. 
Therefore, the vacuum would be again at the middle of the classical Higgs branch where, probably, ${\cal V}$ is positive and the supersymmetry is broken. 

For the remaining cases $N\le k$, there is again no superpotential, and also no constraints like (\ref{constraint}) which forbids the origin of the classical Higgs branch. 
Therefore, the intersection of the Coulomb branch and the Higgs branch would be allowed to be a vacuum state. 
As conjectured in the case $m=0$, the intersection point would correspond to a non-trivial IR fixed point. 
In the case $m\ne 0$, the fixed point would describe ${\cal N}=2$ pure Chern-Simons theory. 

\vspace{5mm}

An interesting point of the analysis shown above is that the supersymmetry breaking occurs in the Higgs branch which is integrated out 
when one obtains the low energy description in terms of Yang-Mill-Chern-Simons theory. 
From this point of view, it is quite natural that the supersymmetry breaking is rather difficult to see in the latter description 
since it occurs in a ``hidden'' part of the theory. 

\vspace{5mm}

It should be noted that the behavior of the quantum D-term potential at infinity of the classical Higgs branch, that is, 
growing toward infinity for the case $m\ne0$, is crucial for the above argument. 
To realize this behavior, the real mass must be the axial mass. 
A vector mass, say $\mu$, would give the behavior of the D-term potential
\begin{equation}
{\cal V} \sim V_D+|(\phi+\mu)q^I|^2+|(\phi+\mu)\widetilde{q}_I|^2. 
\end{equation}
This may have flat directions near infinity, and therefore, the above argument does not lead us to the conclusion that the supersymmetry must be broken. 
Indeed, this seems to be plausible. 
The theory with a vector mass is reduced to ${\cal N}=2$ pure Yang-Mills theory, whose superpotential has runaway directions \cite{deBoer:1997kr}\cite{Aharony:1997bx}.

\vspace{1cm}

\section{Chern-Simons-matter theories of quiver type} \label{quiver type}

\vspace{5mm}

In this section, we consider the vacuum states of ${\cal N}=2$ Chern-Simons theory with the gauge group ${\rm SU}(N_1)\times{\rm SU}(N_2)$ and the level $(+k,-k)$ 
coupled to a number $n_b$ of pairs $(B^n,\widetilde{B}^n)$ $(n=1,\cdots,n_b)$ of bi-fundamental chiral multiplets in the representations $(N_1, \overline{N_2})$ and 
$(\overline{N_1}, N_2)$, respectively. 
The extension of the following arguments to the case with a general $(k_1,k_2)$ will be straightforward. 
Without loss of generality, we assume $N_1\ge N_2$ and $k>0$. 

As in the previous section, we add suitable Yang-Mills terms for the gauge fields, and replace the Chern-Simons terms with the matter chiral multiplets in 
the representations 
\begin{equation}
(N_1,1)^k\oplus (\overline{N_1},1)^k\oplus (1,N_2)^k\oplus (1,\overline{N_2})^k 
\end{equation}
of the gauge group with axial masses specified below. 
We denote them as $Q_1^I, \widetilde{Q}_1^I, Q_2^I$ and $\widetilde{Q}_2^I$, respectively. 
The Lagrangian is 
\begin{eqnarray}
L 
&=& \int d^4\theta \Bigl[ Q_{1I}e^{V_1+m_1\bar{\theta}\theta}Q_1^I+\widetilde{Q}_1^Ie^{-V_1+m_1\bar{\theta}\theta}\widetilde{Q}_{1I}
       +Q_{2I}e^{V_2-m_2\bar{\theta}\theta}Q_2^I+\widetilde{Q}_2^Ie^{-V_2-m_2\bar{\theta}\theta}\widetilde{Q}_{2I}  \nonumber \\
& &  \hspace*{1cm} +\mbox{tr}_2(B_ne^{V_1}B^ne^{-V_2})+\mbox{tr}_1(\widetilde{B}_ne^{V_2}\widetilde{B}^ne^{-V_1}) \Bigr] \nonumber \\
& &  + \int d^2\theta \Bigl[ \mbox{tr}_1(W_1^\alpha W_{1\alpha}) + \mbox{tr}_2(W_2^\alpha W_{2\alpha}) + W_{\rm tree}(B, \widetilde{B}) \Bigr] + \mbox{h.c.} 
   \label{quiver}
\end{eqnarray}
where tr$_1$ and tr$_2$ are the traces for the fundamental representations of SU$(N_1)$ and SU$(N_2)$, respectively. 

\vspace{5mm}

First, let us consider the case $n_b=1$. 
The tree level superpotential $W_{\rm tree}$ would be of the form 
\begin{equation}
W_{\rm tree} = c\,\mbox{tr}_1(B\widetilde{B}B\widetilde{B}). 
\end{equation}
The quadratic term should be absent since we assume the bi-fundamental fields to be massless. 
It was argued in \cite{Gaiotto:2007qi} that, in the context of the Chern-Simons-matter theory, 
$c=0$ would be unstable and flow to a non-zero but finite value due to $1/k$ corrections. 
We assume in the following that $c$ is non-zero and finite. 

\begin{figure}[tbp]
\begin{center}
\includegraphics{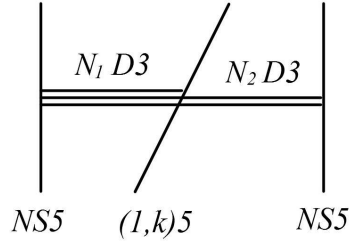}
\caption{The brane configuration realizing ${\cal N}=2$ ${\rm U}(N_1)_k\times{\rm U}(N_2)_{-k}$ Chern-Simons theory coupled to 1 pair of bi-fundamental matters. }
   \label{brane2}
\end{center}
\end{figure}

A similar theory can be realized as the worldvolume theory on D3-branes suspended between two NS5-branes and a $(1,k)$5-brane. 
See Figure \ref{brane2}. 
There are $N_1$ D3-brane segments between the left NS5-brane and the $(1,k)$5-brane, and $N_2$ D3-brane segments between the right NS5-brane and the 
$(1,k)$5-brane. 
The gauge group is ${\rm U}(N_1)\times{\rm U}(N_2)$. 
It is always possible for $2N_2$ D3-brane segments in the above two D3-brane stacks to form longer segments suspended between two NS5-branes which can be 
detached from the $(1,k)$5-brane. 
The remaining $N_1-N_2$ D3-brane segments are still suspended between the left NS5-brane and the $(1,k)$5-brane. 
As in the case for pure Chern-Simons theory, the supersymmetry is broken if $N_1-N_2>k$, according to the s-rule \cite{Hanany:1996ie}. 

We expect that the presence of ${\rm U}(1)$ factors will not be important in the analysis of supersymmetry breaking in this section. 
In terms of the scalar potential, an extra ${\rm U}(1)$ factor provides an additive positive contribution to the D-term potential. 
If the supersymmetry is already broken in the theory without the ${\rm U}(1)$ factor, it is also the case in the theory with the ${\rm U}(1)$ factor. 
In the following, we analyze the pattern of supersymmetry breaking in the theory (\ref{quiver}) which should be compatible with the pattern expected from the argument 
based on the brane configuration in Figure \ref{brane2}. 

\vspace{5mm}

\subsection{Classical vacua}

\vspace{5mm}

Let us start with the analysis of the classical scalar potential ${\cal V}$ of (\ref{quiver}). 
The explicit form of ${\cal V}$ is 
\begin{eqnarray}
{\cal V} 
&=& V_F+V_{D_1}+V_{D_2}+\Bigl| \phi_1b-b\phi_2 \Bigr|^2+\Bigl| \phi_2\widetilde{b}-\widetilde{b}\phi_1 \Bigr|^2 \nonumber \\
& & +\sum_{I=1}^k \Bigl( \bigl| (\phi_1+m_1)q_1^I \bigr|^2+\bigl| (\phi_1-m_1)\widetilde{q}_{1I} \bigr|^2 \Bigr) \nonumber \\
& & +\sum_{I=1}^k \Bigl( \bigl| (\phi_2-m_2)q^I_2 \bigr|^2+\bigl| (\phi_2+m_2)\widetilde{q}_{2I} \bigr|^2 \Bigr). 
\end{eqnarray}
If $m_1=m_2=0$, then there exist flat directions consisting of several branches. 

Consider the case with non-zero $m_1$ and $m_2$. 
The condition $V_F=0$ implies 
\begin{equation}
\widetilde{b}b = 0. 
\end{equation}
The adjoint scalars $\phi_1$ and $\phi_2$ can be diagonalized by a suitable gauge transformation, 
\begin{eqnarray}
\phi_1 
&=& \left[ 
\begin{array}{ccccc}
+m_1\cdot I_{l_1} & & & & \\ 
& -m_1\cdot I_{l_2} & & & \\
& & +m_2\cdot I_{l_3} & & \\
& & & -m_2\cdot I_{l_4} & \\
& & & & \phi_1' 
\end{array}
\right], \\
\phi_2 
&=& \left[ 
\begin{array}{ccccc}
+m_1\cdot I_{l'_1} & & & & \\ 
& -m_1\cdot I_{l'_2} & & & \\
& & +m_2\cdot I_{l'_3} & & \\
& & & -m_2\cdot I_{l'_4} & \\
& & & & \phi_2' 
\end{array}
\right], 
\end{eqnarray}
where $\phi'_1$ and $\phi_2'$ are diagonal matrices whose eigenvalues are not $\pm m_1$ nor $\pm m_2$. 
Then, the flat directions for the mass terms of the fundamental matters are parametrized as 
\begin{equation}
q^I_1 = \left[ 
\begin{array}{c}
0 \\ q^I_{1,\rm flat} \\ 0 \\ 0  \\ 0
\end{array}
\right], \hspace{5mm} \widetilde{q}^I_{1} = \left[ 
\begin{array}{c}
\widetilde{q}^I_{1,\rm flat} \\ 0 \\ 0 \\ 0 \\ 0
\end{array}
\right], \hspace{5mm} q^I_2 = \left[ 
\begin{array}{c}
0 \\ 0 \\ q^I_{2,\rm flat} \\ 0 \\ 0
\end{array}
\right], \hspace{5mm} \widetilde{q}^I_2 = \left[ 
\begin{array}{c}
0 \\ 0 \\ 00 \\ \widetilde{q}^I_{2,\rm flat} \\ 0 
\end{array}
\right], 
\end{equation}
and the flat directions of the bi-fundamental matters are 
\begin{equation}
b = \left[ 
\begin{array}{ccccc}
b_1 & & & & \\ 
& b_2 & & & \\
& & b_3 & & \\
& & & b_4 & \\
& & & & b_5
\end{array}
\right], \hspace{5mm} \widetilde{b} = \left[ 
\begin{array}{ccccc}
\widetilde{b}_1 & & & & \\ 
& \widetilde{b}_2 & & & \\
& & \widetilde{b}_3 & & \\
& & & \widetilde{b}_4 & \\
& & & & \widetilde{b}_5
\end{array}
\right]
\end{equation}

The condition $V_{D_2}=0$ then implies 
\begin{equation}
\left[ \mbox{tr}_1 \left( b_1T^ab_1^\dag \right)-\mbox{tr}_1 \left( \widetilde{b}_1^\dag T^a\widetilde{b}_1 \right) \right]^2 = 0, 
\end{equation}
where $T^a$ are generators of ${\rm su}(l'_1)\subset{\rm su}(N_2)$. 
This can be written as 
\begin{equation}
\left[ \mbox{tr}_1 \left( b_1T^ab_1^\dag \right) \right]^2+\left[ \mbox{tr}_1 \left( \widetilde{b}_1^\dag T^a\widetilde{b}_1 \right) \right]^2
 -2\mbox{tr}_1(\widetilde{b}_1b_1b_1^\dag\widetilde{b}_1^\dag)+\frac 2{l'_1}|b_1|^2|\widetilde{b}_1|^2 = 0. 
\end{equation}
Combined with the F-term condition $\widetilde{b}_1b_1=0$, this implies that either $b_1$ or $\widetilde{b}_1$ vanishes. 
Similarly, it can be shown that for any $i=1,\cdots,5$ either $b_i$ or $\widetilde{b}_i$ vanishes. 

The D-term condition further imposes constraints. 
For example, suppose that $b_1$ and $\widetilde{b}_2$ are non-zero. 
Let $T$ be a generator of ${\rm su}(l'_1+l'_2)\subset {\rm su}(N_2)$ such that 
\begin{equation}
T = \mbox{diag}\left( +l'_2I_{l'_1}, -l'_1I_{l'_2},0,0,0 \right). 
\end{equation}
Then the condition $V_{D_2}=0$ implies 
\begin{equation}
l'_2|b_1|^2+l'_1|\widetilde{b}_2|^2 = 0, 
\end{equation}
concluding that both $b_1$ and $\widetilde{b}_2$ vanish. 
This means that if $b_1$ is non-zero, then $\widetilde{b}_2$ must be zero, and the converse is also true. 
Similar arguments then imply that either $b$ or $\widetilde{b}$ vanish. 

Therefore, the possible flat directions satisfy either $b=0$ or $\widetilde{b}=0$. 
This is the enough information of the classical moduli space of vacua of the theory (\ref{quiver}) for the following discussion of the quantum vacua. 

\vspace{5mm}

\subsection{Quantum vacua}

\vspace{5mm}

Let us consider the quantum effects. 
We first set $m_1=m_2=0$. 

Notice that SU$(N_1)$ gauge fields couple to $k+N_2$ pairs of fundamental chiral multiplets ${\cal Q}^A = (Q^I_1, B)$ and anti-fundamental chiral multiplets 
$\widetilde{\cal Q}^A = (\widetilde{Q}^I_1, \widetilde{B})$, and SU$(N_2)$ gauge fields couple to $k+N_1$ such pairs. 
The dynamics of the former gauge fields will induce a superpotential 
\begin{equation}
W_{\rm np} \propto \Bigl[ Y_1\det(\widetilde{\cal Q}^A{\cal Q}^B) \Bigr]^{\frac1{N_2+k-N_1+1}}+\cdots, 
   \label{quiverNPW}
\end{equation}
if $N_1-1>N_2+k$ is satisfied, while the latter gauge fields will not induce such a superpotential since $N_2-1>N_1+k$ cannot be satisfied for the case $N_1\ge N_2$ which 
we assumed.  
There are other terms in $W_{\rm np}$ depending only on the adjoint scalars. 

The superpotential depends on $Q^I_1$ and $\widetilde{Q}^I_1$ only through the first term in (\ref{quiverNPW}). 
The F-term potential vanishes only if 
\begin{equation}
\det(\widetilde{\cal Q}^A{\cal Q}^B) \to \infty \hspace{5mm} \mbox{or} \hspace{5mm} \mbox{rank}{\cal Q}^A, \mbox{rank}\widetilde{\cal Q}^A \le N_2. 
\end{equation}
The latter possibility is not allowed since in this case $\det(\widetilde{\cal Q}^A{\cal Q}^B)=0$, and therefore, the scalar potential is infinite. 
This then implies that the supersymmetry would be preserved only when there is a runaway direction along which $\det(\widetilde{\cal Q}^A{\cal Q}^B)$ diverges. 

To find whether there really exists such a runaway direction, it is necessary to examine the D-term potential. 
Consider first the region near infinity of the classical flat directions along $q^I_{1,2}$ and $\widetilde{q}^I_{1,2}$ directions. 
Since those fields couple to $b$ and $\widetilde{b}$ only through the gauge fields, $b$ and $\widetilde{b}$ will decouple once $q^I_{1,2}$ and $\widetilde{q}^I_{1,2}$ 
have large vevs so that the corresponding gauge fields become massive and decouple. 
Therefore, the D-term potential in these directions approaches the classical one near infinity along the classical flat directions. 
This is also the case if the vevs of $b$ and $\widetilde{b}$ become large. 
In this case, the gauge group ${\rm SU}(N_1)\times{\rm SU}(N_2)$ would be broken to ${\rm SU}(N_1-N_2-k)$, and all the other light fields are neutral under the remaining 
gauge fields. 
Therefore, no D-term potential will be induced below the energy scale set by the vevs. 
If the vevs are large enough, the D-term potential may be approximated by the classical one. 

Now we turn on small axial masses. 
As was shown in the previous subsection, the possible flat directions in the presence of the masses satisfy $b=0$ or $\widetilde{b}=0$. 
For such a vev, the non-perturbative F-term potential diverges since $\det(\widetilde{\cal Q}^A{\cal Q}^B)$ vanishes. 
In other words, all the classical flat directions in the absence of the masses, which might be compatible with the runaway directions of the F-term potential, are lifted 
by the masses. 
As a result, the vacuum states must correspond to points in the middle of the classical flat directions, at which the F-term potential is non-zero. 
This implies that the supersymmetry is completely broken. 
This conclusion is valid for 
\begin{equation}
N_1-N_2 > k+1, 
\end{equation}
for which the non-perturbative superpotential (\ref{quiverNPW}) is induced. 
This pattern of the supersymmetry breaking is expected to be the same when $m_1$ and $m_2$ become finite but large, suggesting that the analysis above explains the 
supersymmetry breaking of a Chern-Simons-matter theory we would like to discuss. 

It would be possible to argue that the supersymmetry is also broken for the case $N_1-N_2=k+1$, while it is preserved otherwise. 
Note that this pattern is compatible with the one for the ${\rm U}(N_1)\times{\rm U}(N_2)$ theory discussed previously. 

\vspace{5mm}

\subsection{More bi-fundamental matters}

\vspace{5mm}

Next, let us consider the case $n_b\ge 2$. 
The theory with $n_b=2$ can be realized, up to ${\rm U}(1)$ factors, on a D3-brane segments in a brane configuration similar to Figure \ref{brane2}. 
That is, the brane configuration is obtained from the one in Figure \ref{brane2} by compactifying the horizontal direction and identify two NS5-branes, 
as in \cite{Aharony:2008ug}. 
The brane configuration suggests that the supersymmetry is broken when $N_1-N_2>k$ based on the similar argument as for $n_b=1$. 
For the cases $n_b>2$, there are no known brane configurations realizing such theories. 

The argument based on the field theory (\ref{quiver}) is similar to the one in the previous subsection. 
A non-perturbative superpotential is generated when 
\begin{equation}
N_1-1>n_bN_2+k \hspace{5mm} \mbox{and} \hspace{5mm} N_2-1>n_bN_1+k. 
\end{equation}
Under the assumption $N_1\ge N_2$, the second condition is not satisfied, and therefore, a non-perturbative superpotential will be induced only by the dynamics of the 
SU$(N_1)$ gauge group. 
Then, it is suggested that the supersymmetry would be broken completely if 
\begin{equation}
N_1-N_2 > (n_b-1)N_2+k+1. 
\end{equation}
For the case $n_b=2$, this condition is consistent with the one expected from the brane configuration, but certainly much weaker. 
It should be noted that our argument typically provides a sufficient condition for the supersymmetry breaking. 
Even though there is no superpotential, or even though the F-term potential can vanish for some field configurations, there is still a possibility that the D-term potential 
would lift such would-be supersymmetric vacua. 
To obtain a more stringent condition for the supersymmetry breaking, one has to investigate the detailed form of the quantum corrected D-term potential which is, 
however, in general quite difficult for theories with only four supercharges. 

A similar complication would appear when one adds to the theory (\ref{quiver}), or even to (\ref{SQCD}), a large number of (anti-)fundamental matters with {\it vector} 
masses. 
Since such matters would not affect the low energy properties of the theory, the conclusion on the possibility of supersymmetry breaking should not depend on the 
presence of those matters. 
However, if the number of matters in the (anti-)fundamental representation is too large, the non-perturbative superpotential would not be induced. 
In such cases, the only possibility would be that the supersymmetry is broken by the D-term potential, which would be very difficult to show explicitly.

\vspace{2cm}

{\bf \Large Acknowledgements}

\vspace{5mm}

I would like to thank Soo-Jong Rey and Kimyeong Lee for valuable discussions. 
This work was supported by the BK21 program of the Ministry of Education, Science and Technology, 
National Science Foundation of Korea Grants 0429-20100161, R01-2008-000-10656-0, 
2005-084-C00003, 2009-008-0372 and EU-FP Marie Curie Research 
\& Training Network HPRN-CT-2006-035863 (2009-06318).

\end{document}